\documentclass{aa}
\newcommand {\hii}{H\,{\sc ii}} 
\newcommand {\hei}{He\,{\sc i}} 
\newcommand {\heii}{He\,{\sc ii}} 
\newcommand {\kms}{\relax \ifmmode {\,\rm km\,s}^{-1}\else \,km~s$^{-1}$\fi}
\newcommand {\ha}{H$\alpha$}
\newcommand {\hb}{H$\beta$}
\newcommand {\hg}{H$\gamma$}
\newcommand {\hd}{H$\delta$}
\newcommand {\he}{H$\epsilon$}
\newcommand {\oiii}{[O\,{\sc iii}]}
\newcommand {\oii}{[O\,{\sc ii}]}
\newcommand {\oi}{[O\,{\sc i}]}
\newcommand {\nii}{[N\,{\sc ii}]}
\newcommand {\sii}{[S\,{\sc ii}]}
\newcommand {\siii}{[S\,{\sc iii}]}
\newcommand {\neiii}{[Ne\,{\sc iii}]}
\newcommand {\ariv}{[Ar\,{\sc iv}]}

\usepackage{graphicx}
\usepackage{txfonts}

\begin{document}

   \title{VLT observations of the highly ionized 
nebula around Brey2\thanks{Based on observations collected at 
the European Southern Observatory, Cerro Paranal, Chile (ESO No. 
68.C-0238(A,B)).}}

   \subtitle{}

   \author{Y. Naz\'e\thanks{Research Fellow FNRS (Belgium)}\inst{1}
	  G. Rauw\thanks{Research Associate FNRS (Belgium)} \inst{1}
          J. Manfroid\thanks{Research Director FNRS (Belgium)}  \inst{1}
	  Y.-H. Chu \inst{2} \and J.-M. Vreux \inst{1}}

   \offprints{Y. Naz\'e \email{naze@astro.ulg.ac.be}}

   \institute{Institut d'Astrophysique et de G\'eophysique;
	Universit\'e de Li\`ege;
	All\'ee du 6 Ao\^ut 17, Bat. B5c;
	B 4000 - Li\`ege;
	Belgium 
\and Astronomy Department;
University of Illinois at Urbana-Champaign;
	1002 West Green Street;
	 Urbana, IL 61801;
	USA
             }
\authorrunning{Naz\'e et al.}
   \date{}

   \abstract{ 

We present the first high resolution \heii\ $\lambda$4686 images 
of the high excitation nebula around the WR star Brey 2 in the LMC.
This nebula presents a striking morphology: a small arc-like feature 
some 3.6~pc in radius is particularly prominent in the \heii\ 
$\lambda$4686 line. We further discover a previously unknown faint
\heii\ emission that extends over an area of 22$\times$17~pc$^2$. 
An even fainter \heii\ emission is apparently associated
with the interstellar bubble blown by the progenitor of Brey2. 
The total \heii\ flux corresponds to an ionizing flux of 4$\times$10$^{47}$ photons s$^{-1}$. \ha, \oiii, and \hei\ 
$\lambda$5876 images and long-slit spectra are also examined in this 
letter, enabling us to investigate the detailed physical properties 
at various locations of the nebula.

   }

   \maketitle
\section{Introduction}

Except for Planetary Nebulae, high excitation features, such as 
nebular \heii\ emission, are not expected in \hii\ regions.
Indeed, normal O stars do not emit enough hard UV photons
to produce detectable nebular \heii\ emission.
However, in the last two decades, seven 
objects of the Local Group were found to harbour such nebular \heii\ 
emission (Garnett et al.  \cite{gar91}). Wolf-Rayet (WR) stars 
are the ionizing sources of five of them. These stars are hot, 
evolved massive stars that possess strong winds, but until then 
they were generally believed to be unable to excite such highly 
ionized nebulae. Nebular \heii\ emission was also 
discovered farther away, in star-forming galaxies (e.g. I Zw 18, 
see  French \cite{fre}) and WR stars were again thought 
to be responsible for it (Schaerer \cite{sch}). This type of emission 
thus seems intimately linked to WR stars in most cases, and the 
analysis of these peculiar highly ionized objects can help us  
put tighter constraints on the poorly known extreme UV ionizing fluxes 
of these stars (Crowther \cite{cro}). Only the \heii\ nebulae 
of the Local Group are situated close enough to enable a detailed study.

We study here the nebula associated with Brey2 (or BAT99-2 in Breysacher 
et al. \cite{bat}), a WN2b(h) star in the Large Magellanic Cloud 
(Foellmi et al. \cite{foe}).
This WR star has blown a small bubble whose expansion velocity
reaches 16\kms\ (Chu et al. \cite{chu}). Moderate chemical
enrichment was detected in this bubble (Garnett \& Chu 
\cite{gar}). Using this result, Chu et al. (\cite{chu}) suggested 
that the circumstellar bubble was probably currently merging with the
interstellar bubble blown by the progenitor of the WR. 
Nebular \heii\ emission near Brey2 was first detected by Pakull 
(\cite{pak}). Imaging of the nebula in the \heii\ lines is  
crucial in establishing the distribution of the emission in detail,
in order to evaluate the total flux, and to better 
understand the origin of the ionization. To date, 
only low resolution, low signal/noise \heii\ images exist (Melnick \& 
Heydari-Malayeri \cite{mel}). 
Thanks to observations made with the Very Large Telescope (VLT), we 
present here the first high resolution \heii\ $\lambda$4686 images 
and a detailed 
spectral analysis of this peculiar highly ionized nebula.

\section{Observations}

We obtained CCD images of Brey2 and its associated nebula with the FORS
instrument installed on the 8~m VLT-UT3 in 2002 January. The images 
were taken through seven filters (\ha, \oiii, \hei\ $\lambda$5876, \heii\ $\lambda$4686, 
plus three continuum filters centered on 4850, 5300, and 6665 \AA) 
for exposure times of 3$\times$100s, 3$\times$100s, 9$\times$400s, 
9$\times$400s, 
9$\times$60s, 9$\times$60s, and 3$\times$100s, respectively. The 
seeing was $\sim$1\arcsec. The data were reduced with {\sc iraf} 
using standard methods for overscan and bias subtraction 
and flatfielding. 
Stellar sources were removed using the photometric and astrometric
information obtained from the continuum images (details are reported by
Naz\'e et al., in preparation).
The few remaining faint stars were either removed individually or not
considered for flux determinations. Fig. \ref{br2ha} presents a three color
image of the whole field; while Fig. \ref{mosaic} shows a close-up on Brey2
in the four nebular filters.

During the same observing run, we also obtained long-slit spectra of 
Brey2 with the same instrument.  We used the 600B and 600V grisms to 
obtain a blue spectrum covering 3700-5600 \AA\ ($R\sim800$) and a red 
spectrum covering 4500-6850\AA\ ($R\sim1000$), respectively. The 
1.3\arcsec$\times$6.8$'$ slit was tilted by 45$^{\circ}$ and centered 
on Brey2. The spatial resolution was $\sim$1.2\arcsec\ and the spectral 
resolution, as mesured from the FWHM of the calibration lines, 7\AA. 
The spectra were reduced and calibrated in a standard way using {\sc 
iraf}. For flux calibration, we observed several standard stars from 
Oke (\cite{oke}) and used the mean atmospheric extinction 
coefficients for CTIO reduced by 15\%. Sky subtraction was done using 
a small region of the spectra where the nebular emission is the
lowest. Only a few residuals remained for the brighter sky lines 
(e.g. \oi\ 5577\AA).

\section{The high excitation nebula surrounding Brey2}

\subsection{Morphology}

   \begin{figure}
   \centering
   \caption{Color image of the Brey2 field. Red, green and blue 
correspond to continuum subtracted \ha, \oiii, and \heii\ images, 
respectively. The different regions used for spectral analysis 
are marked by a solid line. Features discussed in the text 
are labelled.}
              \label{br2ha}
    \end{figure}

   \begin{figure}
   \centering
   \caption{FORS continuum subtracted \ha, \oiii, \heii, 
and \hei\ images of Brey2 and its close surroundings. The images 
are 200\arcsec$\times$200\arcsec. A white cross indicates the 
WR star's position.}
              \label{mosaic}
    \end{figure}

As seen in Fig. \ref{br2ha}, the nebula surrounding 
Brey2 is quite complex. The bubble blown by the WR star 
is visible to the NW of the star as a small, quite bright 
arc(and arrow)-like structure, of external radius $\sim$15\arcsec. 
To the south of the star, a large shell, 
$\sim$100$''$ in radius, may in fact be the interstellar bubble 
blown by the progenitor of Brey2 (Garnett \& Chu \cite{gar}).  
Though with a low S/N, the \hei\ image appears well correlated with 
\ha, at least for the brightest features. But the \oiii\ 
image seems less filamentary than the \ha\ image at the SNR's position 
(see below) and at the SE of the field-of-view. It also shows several 
additional features compared to the \ha\ image. 
All \ha\ features on the eastern side of the nebula possess a 
lower \oiii/\ha\ ratio than the central nebulosities surrounding Brey2.
BSDL 49 (Bica et al. \cite{bic}), 
an oval-shaped structure 160\arcsec\ west of Brey2 and 
140\arcsec$\times$100\arcsec\ in size, presents an extremely low 
\oiii/\ha\ ratio.  These low ratios suggest that the regions are not 
ionized by early O or WR stars but most probably by cooler stars, 
e.g. late O or early B stars.

In contrast, the \heii\ image appears strikingly different (see Fig.
\ref{mosaic}). The 
arc-like structure emits strongly in \heii, and a fainter halo  
$\sim$90\arcsec$\times$70\arcsec\ in size surrounds the star.  
These \heii\ emissions appear well centered on Brey2. The shell 
directly south of the star also emits \heii, but rather faintly. 
This shell may represent the limit of the \heii\ Str\"omgren sphere.
Since the density south of Brey2 appears much lower than in the other 
directions, some \heii\ ionizing photons could escape the direct 
surroundings of Brey2 and ionize this shell. Finally, note that the 
very faint halo seems to extend to the bright \ha\ regions visible
at the  lower left of Fig. \ref{br2ha}. However, since we do not have 
spectra of this region,
we can not confirm that the nebular emission actually comes from \heii, 
and not from other nebular lines, e.g. \ariv.

We used the calibrated long-slit spectra to get the ADU-flux 
conversion factor of the images. The total \heii\ emission arc$+$halo 
is about 5$\times$10$^{-13}$ erg cm$^{-2}$ s$^{-1}$, or a luminosity 
of 3 $\times$10$^{35}$ erg s$^{-1}$ at the distance of the LMC (50~kpc),
and after a reddening correction of $A_{{\rm V}}=0.7\,{\rm mag}$ 
(see \S 3.2). This
corresponds to a flux of 4$\times$10$^{47}$ He$^+$ ionizing photons s$^{-1}$.
Comparing to the models of Smith et al. (\cite{smi}), this flux 
is compatible with a 90-100~kK WR star, consistent with the WN2 
classification of Brey2.

To the NW of Brey2, a SNR was diagnosed by its high-velocity motions 
and confirmed through nonthermal radio emission and its high \sii/\ha\ ratio 
(Chu et al. \cite{chu}). Between BSDL 49 and the features associated 
with Brey2, filamentary structures, characteristic of SNRs (Chen et al. 
\cite{che}), prove the presence of this SNR candidate. Two 
rather bright \ha\ filaments seem to delineate the SNR to the north 
and south. They form a region some $\sim$50\arcsec\ in radius. Our long-slit 
spectra taken at that position show clear indications of an enhanced 
\sii/\ha\ ratio, typical of SNRs (see below), and a lower \oiii/\ha\ ratio.

\subsection{Spectrophotometry}

   \begin{figure}
   \centering
   \includegraphics[width=9cm]{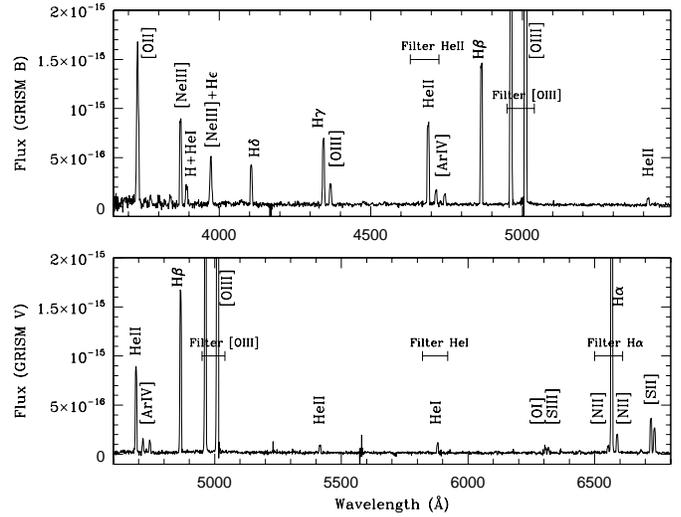}
   \caption{FORS calibrated, sky-subtracted blue (upper) and red 
(lower) spectra of the arc near Brey2. The lines analysed in Table 
\ref{linerat} are labelled, and a solid line shows the bandpass of
the imaging filters. Note that the \heii\ filter also includes 
the \ariv\ lines, whilst the \ha\ filter also contains the \nii\ lines.}
              \label{arc}
    \end{figure}

The Balmer decrement \ha/\hb\ was used to derive the interstellar 
extinction, assuming the theoretical case B decrement at $T=15000\,{\rm K}$
and $30000\,{\rm K}$ (Storey \& Hummer \cite{sto}). Some faint stars sometimes 
affected the nebular spectra. 
In such cases, we correct the measured emission line strengths by a 2 
\AA\ equivalent width for the Balmer absorptions (McCall et al. 
\cite{mcc}) before estimating the reddening. 
To deredden the line ratios, we used the extinction law from Cardelli 
et al. (\cite{car}) with $R_{{\rm V}}=3.1$. 
Spectra were extracted from five regions of interest for detailed 
analysis (Fig. \ref{br2ha}): the first one includes the arrow of 
the arc-like structure (12\arcsec\ long without the star); the second 
and third ones explore the eastern and western \heii\ halos (34\arcsec\ 
and 22\arcsec\ long, respectively); the fourth examines the eastern 
part of the large southern shell (40\arcsec\ long); and the fifth 
investigates the properties of the SNR candidate situated to the 
NW of the star (44\arcsec\ long). The extinction-corrected line 
ratios for these regions are listed in Table \ref{linerat}.

Using the {\sc nebular} package under {\sc iraf}
(Shaw \& Dufour \cite{sha}), we then derived temperatures, densities
and abundances in these regions (see Table \ref{abund}). To get Helium 
abundances, we used the emissivities computed by Benjamin et al.
(\cite{ben}) for \hei\ and by Storey \& Hummer (\cite{sto}) for
H and \heii. The presence of \heii\ suggests that higher ionization
states than O$^{2+}$ exist as well. This can actually explain the slightly 
smaller \oiii/\ha\ ratio of the arc-like structure compared to 
its direct surroundings. Following Garnett \& Chu (\cite{gar}), 
we derived the total oxygen abundance using : \\
$\frac{N({\rm O})}{N({\rm H})}=\frac{N({\rm O}^{0+})+N({\rm O}^+)+N({\rm O}^{2+})}{N({\rm H}^+)} \times \left( 1+\frac{N({\rm He}^{2+})}{N({\rm He}^+)}\right)$\\
Compared to the LMC abundances (Russell \& Dopita \cite{rus}), no 
sign of enrichment in helium can be detected for the \heii\ nebula. 
But its oxygen abundance is slightly lower (max 35\%), and its N/O ratio
slightly higher (min 30\%) than the LMC's trend. This is compatible 
with what might be expected for a nebula enriched by a stellar wind. 
Using a closer background for analysing the spectrum of the arc does not
change this conclusion. On the other hand, the 
SNR shows rather low helium and oxygen abundances, but a normal N/O 
ratio. 

\begin{table*}[htb]
\begin{center}
\caption{Dereddened line ratios with respect to \hb=100.0 (see text 
for details). The extinction was calculated using the \ha/\hb\ ratio.
Uncertainties, estimated from the signal/noise 
in the lines and the calibration errors, are given in parentheses.
\label{linerat}} 
\begin{tabular}{l c c c c c c } 
\hline\hline
 Line & $A(\lambda)/A_{{\rm V}}$& Arc & Halo E & Halo W & S shell & SNR\\ 
\hline
\oii\ 3727& 1.54&132 (14) & 268 (28) &225 (23) & 344 (36) &853 (89)\\
\neiii\ 3868&  1.50&69 (7) &  96 (10) & 93 (9) &  85 (8) & 84 (8) \\
H8 + \hei\ 3889& 1.50& 17 (2) &  19 (2) & 19 (5) &  17 (2) & \\
\neiii\ + \he\ 3967 & 1.47& 38 (4) & 42 (4) & 44 (4) & 31 (3) & 23 (2) \\
\hd& 1.43& 30 (3) &  30 (3) & 30 (3) & 28 (3) & 27 (3) \\
\hg& 1.35& 50 (4) &  52 (4) & 51 (4) & 50 (4) & 52 (4) \\
\oiii\ 4363& 1.34& 16  (1) &  16 (1) & 16 (1) &  13 (1) & 28 (2) \\
\hei\ 4471& 1.30& &  & 3.6  (0.3) &  &\\
\heii\ 4686& 1.22& 56 (4) &  21 (01) & 24 (2) &  8.1 (0.6) &\\
\ariv\ 4711$^a$& 1.21& 8.2 (0.6) &  & 3.8 (0.3) &  & \\
\ariv\ 4740& 1.20& 6.9 (0.5) &  & 3.8 (0.3) &  &\\
\hb & 1.16 & 100. & 100.& 100.& 100.&100. \\
\oiii\ 4959& 1.13&234 (17) & 313 (22) &318 (23) & 272 (20) &165 (12)\\
\oiii\ 5007& 1.12&691 (49) & 945 (67) &944 (67) & 804 (58) &489 (35)\\
\heii\ 5412& 1.02& 4.5 (0.4) &  & 2.1  (0.2) &  &\\
\hei\ 5876 & 0.93& 5.2 (0.5) &  10 (1) & 9.6 (0.8) &  11  (1) & 9.1  (0.8)\\ 
\oi\ 6300  & 0.86& 3.1 (0.3) &  4.9 (0.5)& 7.3 (0.7) &  & 25 (2)\\
\siii\ 6312& 0.86& 2.6 (0.3) &  3.7 (0.4)& 3.0 (0.3) &  &\\
\oi\ 6364& 0.85&& & 2.4 (0.2) &  & 8.8 (0.8)\\
\nii\ 6548 & 0.82& 3.4 (0.3) &  7.9 (0.8) & 5.5 (0.6) & 10 (1) & 17 (2)\\
\ha$^b$& 0.82&279 (28) & 279 (28) &279 (28) & 279 (28) &270 (27)\\
\nii\ 6583& 0.82&10 (1) & 22 (2) & 19 (2) & 30 (3) & 54 (5)\\
\hei\ 6678& 0.80&1.7 (0.2)&2.7 (0.3)&2.9 (0.3)&4.3 (0.5)&\\
\sii\ 6716& 0.79&16 (2) & 31 (3) & 29 (3) & 40 (4) & 93 (10)\\
\sii\ 6731& 0.79&12 (1) & 22 (4) & 21 (2) & 29 (3) & 68 (7)\\
$F$(\hb) (erg cm$^{-2}$ s$^{-1}$)& &1.16e-14& 1.30e-14& 1.93e-14& 1.25e-14& 7.47e-15\\
$A_{{\rm V}}$(mag) && 0.65 (0.22)& 0.70 (0.22)& 0.72 (0.22)& 0.62 (0.22)& 1.06 (0.22)\\
\hline
\end{tabular}
\end{center}
$^a$ The small pollution due to \hei\ 4713 was corrected using the  
strength of \hei\ 5876 and the theoretical \hei\ ratios from Benjamin 
et al. (\cite{ben}).\\
$^b$ Pollution due to \heii\ 6560 negligible.\\
\end{table*}

\begin{table*}[htb]
\begin{center}
\caption{Derived physical properties of the 5 regions. The abundances 
were calculated assuming $n_{{\rm e}}=100\,{\rm cm}^{-3}$. If several lines of the 
same ion exist, the abundance shown here is an average of the abundances 
derived for each line.  \label{abund}} 
\begin{tabular}{l c c c c c } 
\hline\hline
 Prop. & Arc & Halo E & Halo W & S shell & SNR\\ 
\hline
$T_{{\rm e}}$\oiii\ (K) &16100$\pm$750& 14300$\pm$600& 14100$\pm$650& 14000$\pm$600& 29800$\pm$2700\\
$n_{{\rm e}}$\sii\ (cm$^{-3}$) & $<$350& $<$310& $<310$& $<$280& $<370$\\
$n_{{\rm e}}$\ariv\ (cm$^{-3}$) & 950-4000& & 3200-7000& & \\
\hline
He$^+$/H$^+\times 10^{2}$ & 4.44$\pm$0.30& 7.78$\pm$0.53& 7.70$\pm$0.40& 10.3$\pm$0.7& 7.17$\pm$0.62\\
He$^{2+}$/H$^+\times 10^{2}$ & 4.91$\pm$0.26& 1.80$\pm$0.13& 2.19$\pm$0.12& 0.70$\pm$0.05& \\
$\rightarrow$ He/H$\times 10^{2}$ & 9.35$\pm$0.40 & 9.58$\pm$0.55& 9.89$\pm$0.42 & 11.0$\pm$0.7 & 7.17$\pm$0.62\\
O$^{0+}$/H$^+\times 10^{6}$ & 1.26$\pm$0.12& 2.80$\pm$0.27& 4.47$\pm$0.31& & 2.53$\pm$0.17\\
O$^+$/H$^+\times 10^{5}$ & 0.90$\pm$0.09& 2.64$\pm$0.27& 2.32$\pm$0.24& 3.63$\pm$0.38& 3.25$\pm$0.34$^b$\\
O$^{2+}$/H$^+\times 10^{4}$ & 0.61$\pm$0.03& 1.12$\pm$0.05& 1.17$\pm$0.05& 1.02$\pm$0.05& 0.13$\pm$0.01\\
$\rightarrow$ O/H$\times 10^{4}$ & 1.51$\pm$0.09& 1.73$\pm$0.08& 1.86$\pm$0.08& 1.48$\pm$0.06& 0.48$\pm$0.01$^b$\\
N$^+$/H$^+\times 10^{6}$ & 0.69$\pm$0.05& 1.95$\pm$0.14& 1.54$\pm$0.11& 2.64$\pm$0.19& 1.38$\pm$0.10\\
($\rightarrow$ N/O$\times 10^{2}$)$^a$ & 7.66$\pm$0.98& 7.39$\pm$0.93& 6.64$\pm$0.84& 7.29$\pm$0.92& 4.24$\pm$0.53$^b$\\
S$^+$/H$^+\times 10^{7}$ & 2.44$\pm$0.19& 5.70$\pm$0.45& 5.51$\pm$0.41&7.72$\pm$0.59& 6.07$\pm$0.44\\
S$^{2+}$/H$^+\times 10^{6}$ & 1.17$\pm$0.11& 2.36$\pm$0.23& 2.01$\pm$0.19& & \\
Ar$^{3+}$/H$^+\times 10^{7}$ & 4.31$\pm$0.23& & 2.93$\pm$0.16& & \\
Ne$^{2+}$/H$^+\times 10^{5}$ & 1.55$\pm$0.15& 3.01$\pm$0.30& 3.04$\pm$0.30&2.85$\pm$0.28 & 0.50$\pm$0.05 \\
\hline
\end{tabular}
\end{center}
$^a$ Assuming $N({\rm N}^+)/N({\rm O}^+)=N({\rm N})/N({\rm O})$\\
$^b$ To consider with caution : the calculation of the \oi\ abundance was made at 20 000 K.\\
\end{table*}

\section{Conclusion}

We have presented the first high resolution images of the nebula around
Brey2, one of the
7 \heii\ nebulae of the Local Group. \heii\ is particularly emitted
by a small arc-like structure, some $\sim$15\arcsec\ in radius, that 
surrounds the star. This arc corresponds to the bubble blown 
by the WR star. Fainter \heii\ emission is present in a surrounding 
halo, and still fainter \heii\ emission is detected in the large 
interstellar bubble blown by the progenitor of Brey2 towards the 
south of the star. We derived an He$^+$ ionizing flux of 4$\times$10$^{47}$ 
photons s$^{-1}$, which appears consistent with the latest theoretical 
models for very early WN stars.
We also confirm the results of Garnett \& Chu (\cite{gar}) regarding 
the composition of the nebula: with a larger N/O ratio and a slightly 
smaller O abundance, the small nebula associated with Brey2 shows only 
modest enrichment by stellar winds.
We have also identified the SNR in the FORS images, and have analysed its
spectrum. The bright - thus dense - \ha\ region situated between Brey2 and 
the SNR may be the result of the compression of the ISM between these 
two expanding structures. 

\begin{acknowledgements}
We acknowledge support from the PRODEX XMM-OM and Integral Projects 
and through contracts P4/05 and P5/36 `P\^ole d'Attraction Interuniversitaire' 
(Belgium).
\end{acknowledgements}

\end{document}